\documentclass[aps,prl,twocolumn,groupedaddress,showpacs]{revtex4}

\def\ket#1{\mathinner{|{#1}\rangle}}

\usepackage{graphicx}
\begin{document}

\title{Influence of antisymmetric exchange interaction on quantum tunneling of magnetization in a dimeric molecular magnet $\rm\bf Mn_6$}

\author{S. Bahr$^1$, C. J. Milios$^2$, L. F. Jones$^2$ ,E. K. Brechin$^2$ ,V.Mosser$^3$, and W. Wernsdorfer$^1$}

\affiliation{
$^1$Institut N\'eel, associ\'e \`a l'UJF, CNRS, BP 166, 38042 Grenoble Cedex 9, France\\
$^2$School of Chemistry, The University of Edinburgh, West Mains Road, Edinburgh, 0020EH9 3JJ, U.K.\\
$^3$Itron France, 76 avenue Pierre Brossolette, 92240 Malakoff , France}

\date{\today}

\begin{abstract}
We present magnetization measurements on the single molecule magnet $\rm Mn_6$ 
revealing various tunnel transitions inconsistent with a giant spin description.
We propose a dimeric model of the molecule with two coupled spins $S=6$ , which 
involves crystal field anisotropy, symmetric Heisenberg exchange interaction and antisymmetric Dzyaloshinskii-Moriya exchange interaction.
We show that this simplified model of the molecule explains the experimentally observed 
tunnel transitions and that the antisymmetric exchange interaction between the spins 
gives rise to tunneling processes between spin states belonging to different spin multiplets.
\end{abstract}


\maketitle

Single-Molecule Magnets (SMMs) have been studied intensively in recent years 
because of the unique  crossover between classical an quantum 
physics \cite{novak:1995,friedman:1996,thomas:1996,gatteschi:2003}.
The macroscopic observation of quantum phenomena such as tunneling between 
different spin states or quantum interference between tunneling paths give the possibility 
of studying in detail the quantum mechanical laws in nanoscale molecular systems and 
also might provide substantial information concerning the implementation of spin based solid state 
qubits \cite{wernsdorfer:science1999,leuenberger:2001,troiani:2005,ardavan:2007}.

During the last ten years the spin system of SMMs has mainly been described by a single 
macroscopic spin and the associated tunneling processes were transitions inside a 
multiplet with total spin S, i.e. transitions that conserve the total spin S of the molecule \cite{caneschi:jacs1991,sessoli:jacs1993,barra:1996}.
Recent developments in the field of molecular magnetism go beyond this giant-spin 
approximation \cite{petukhov:2004,barco:2004,carretta:2007prl,barco:nature2008,carretta:prl2008}.
When describing the molecule as an object composed of several superexchange-coupled 
spins $s_i$ the total spin S of the molecule is not fixed, but several multiplets with different total 
spin $S$ appear and as a consequence the allowed tunnel transitions and relaxation paths of the spin system increase considerably.
The associated tunnel processes between different spin states in this multi-spin description 
do not need to conserve the total spin $S$ of the molecule. 
Recently \emph{Carretta et al.} showed evidence of this quantum superposition of spin 
states with different total spin length in the molecule magnet $\rm Cr_7 Ni$ by inelastic neutron scattering (INS) \cite{carretta:2007prl}.
In fact, when introducing an antisymmetric exchange coupling 
(Dzyaloshinskii-Moriya interaction) between the spins $s_i$ that compose 
the molecule, the superposition of a symmetric and an antisymmetric spin state
 becomes possible. The associated tunneling process and quantum interference effects 
 of different tunneling paths have been observed recently in a $\rm Mn_{12}$-based molecular wheel \cite{barco:nature2008}.

In this Brief Report we report the observation of quantum tunneling between spin states 
with different total spin S in a Mn-based SMM having presently the highest anisotropy barrier of 89~K \cite{Milios_JACS2}.
A theoretical model is proposed that describes the molecule as an 
exchange-coupled system of two separated spins $s_i=6$. The experimentally 
obtained tunnel splittings that uses the Landau-Zener method of various symmetric 
and antisymmetric tunnel transitions are compared to this theoretical model.

\begin{figure}[b]
\includegraphics[width=3.4 in]{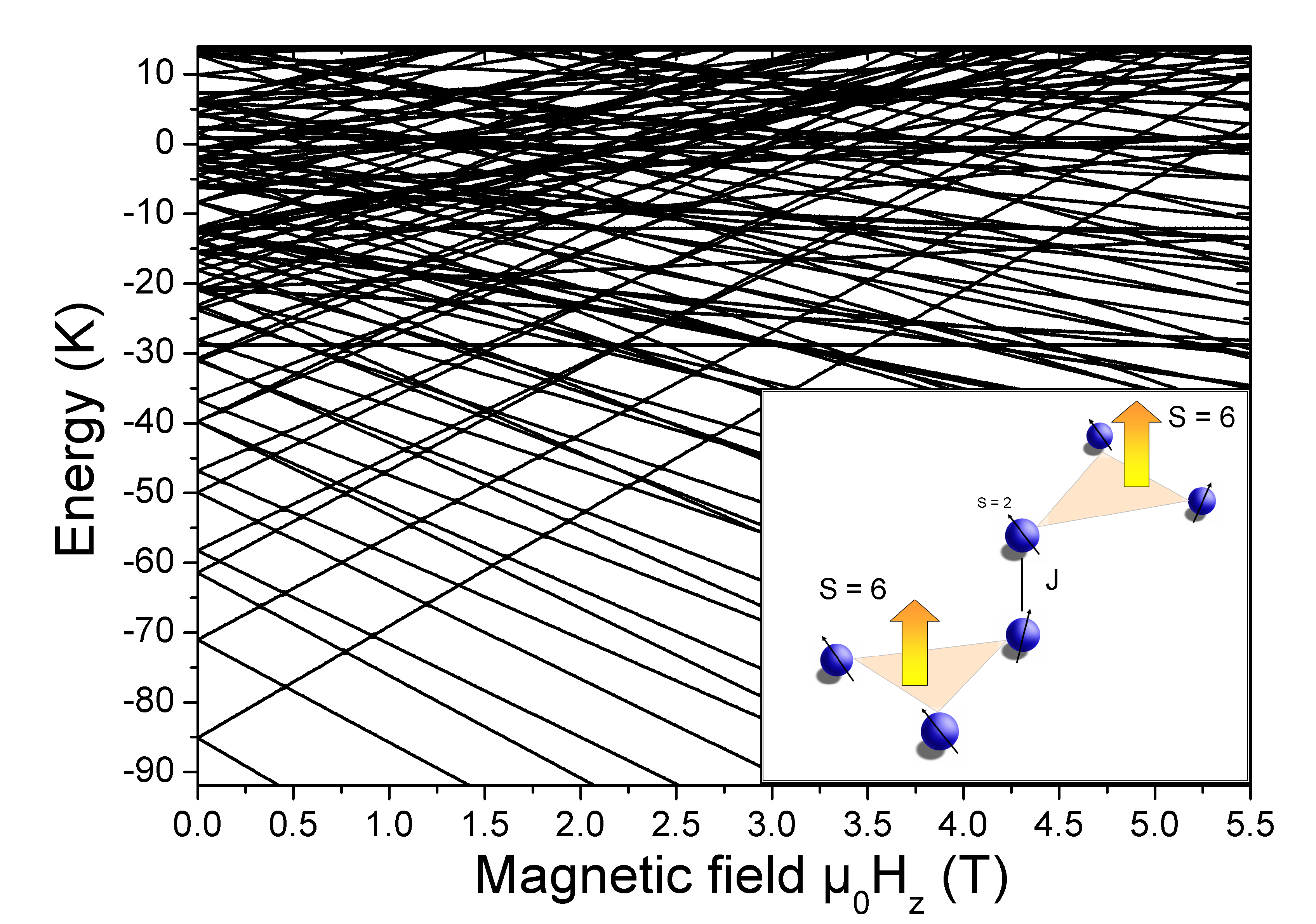}
\caption{\label{fig1} (color online)
Zeeman diagram of the dimeric molecule $\rm Mn_6$ using the longitudinal 
anisotropy constant $D=1.28$~K, an isotropic Heisenberg exchange interaction $J=0.8$~K and $g=1.99$.
Due to the exchange interaction some excited spin multiplets are located 
only a few Kelvins above the ground state.
The inset shows a simplified model of the magnetic core of the $\rm Mn_6$ molecule. 
Two ferromagnetically coupled spin triangles, each having a total spin $S=6$, form the dimeric molecule.
}
\end{figure}

The SMM has the chemical formula [$Mn^{III}_6 O_2 (Et-sao)_6 (O_2 CPh(Me)_2)_2 (EtOH)_6$] 
and will be called briefly $Mn_6$ \cite{Milios_JACS2}.
The six Mn atoms, each having a spin $s_i=2$, form the core of the molecule and they 
are strongly superexchange coupled and act as a macroscopic spin $S=12$ at low temperature.
Recent work by \emph{Carretta et al.} employing INS shows evidence of very low lying 
excited spin multiplets in $\rm Mn_6$ resulting in the breakdown of the giant spin model \cite{carretta:prl2008}.
In contrast to \emph{Carretta et al.} we propose to describe the molecule by two 
superexchange coupled spin triangles, each of them being described by a rigid total spin $S=6$ (see inset of Fig.~1).
This molecular dimer description is in very good agreement with the INS 
measurements and simulations shown by \emph{Carretta et al.}
This simplified model reproduces very well the low-lying spin multiplets and 
gives the advantage of a small Hilbert and parameter space compared to the description of the $\rm Mn_6$ molecule in Ref. \cite{carretta:prl2008}.

Each of the two ferromagnetically coupled spins of the molecular dimer $S_1=S_2=6$ can be described by the spin Hamiltonian :
\begin{eqnarray}
    {\cal H}_i =  -D (S_i^z)^2 + \hat{\cal O}(4) - g \mu_{\rm B} \mu_0 \vec{S}_i\cdot\vec{H}
\end{eqnarray}
where $S_i^{x}$, $S_i^{y}$ and $S_i^{z}$ are the vector components of the $i$th spin operator, 
$g=1.99$ is the gyromagnetic factor and $\mu_{\rm B}$ is the Bohr magneton. \cite{Milios_JACS2}
The first term describes the uniaxial anisotropy of the molecule with longitudinal 
anisotropy parameter $D$ and the second term contains fourth order crystal field anisotropy terms.
The last term is the Zeeman interaction of the spin $\vec{S}_i$ with an external magnetic field $\vec{H}$.

The exchange interaction of the two halves of the molecule can be described by
\begin{eqnarray}
        {\cal H}_{\rm ex} =  -J \vec{S_1}\cdot\vec{S_2} + \vec{D}_{12} \cdot (\vec{S_1}\times\vec{S_2})
\end{eqnarray}
where the first term describes the isotropic Heisenberg exchange interaction with 
exchange constant $J$ and the second term is an antisymmetric Dzyaloshinskii-Moriya interaction between the two spins.

Exact diagonalization of the total spin Hamiltonian ${\cal H} = {\cal H}_1+ {\cal H}_2 + {\cal H}_{\rm ex}$ leads to the energy spectrum shown in figure 1.
The lowest lying spin states belong to the $S=12$ multiplet.
Due to the ferromagnetic exchange the first excited spin multiplet $\ket{S=11,M_S=\pm 11}$ 
is located at about 25~K above the ground state doublet $\ket{S=12,M_S=\pm 12}$ in zero magnetic field.

In the following we will discuss the different level crossings not in the eigenbasis of the total spin of the molecule $\ket{S,M_S}$.
As the total spin of the molecule may fluctuate we chose the more convenient eigenbasis 
of the two single spins of the molecule $\ket{S_1,m_1}\otimes\ket{S_2,m_2} \equiv \ket{m_1,m_2}$. 
The groundstate doublet can be expressed as $\ket{12,\pm 12} \equiv \ket{\pm 6,\pm 6}$ 
and the first doublet of the first excited multiplet reads $\ket{11,\pm 11} \equiv \frac{1}{\sqrt{2}}(\ket{\pm 6, \pm 5}-\ket{\pm 5, \pm 6})$.
The lowest lying spin eigenstates belonging to the ground state multiplet $S=12$ are 
symmetric in respect to a permutation of the two spins in the product base 
$\ket{m_1}\otimes\ket{m_2}$, whereas the eigenstates of the $S=11$ multiplet are antisymmetric.
When we look at the probability to tunnel from one spin state to another, we see 
immediately that most of the terms in the Hamiltonian ${\cal H}$ are symmetric and therefore only provide coupling between symmetric spin states. 
The only antisymmetric term in the Hamiltonian is the Dzyaloshinskii-Moriya exchange 
interaction, and as a consequence this term can provide a coupling between a symmetric 
and an antisymmetric spin state, i.e. this term couples spin states of the ground state multiplet $S=12$ and the first excited multiplet $S=11$.

\begin{figure}[t]
\includegraphics[width=3.4 in]{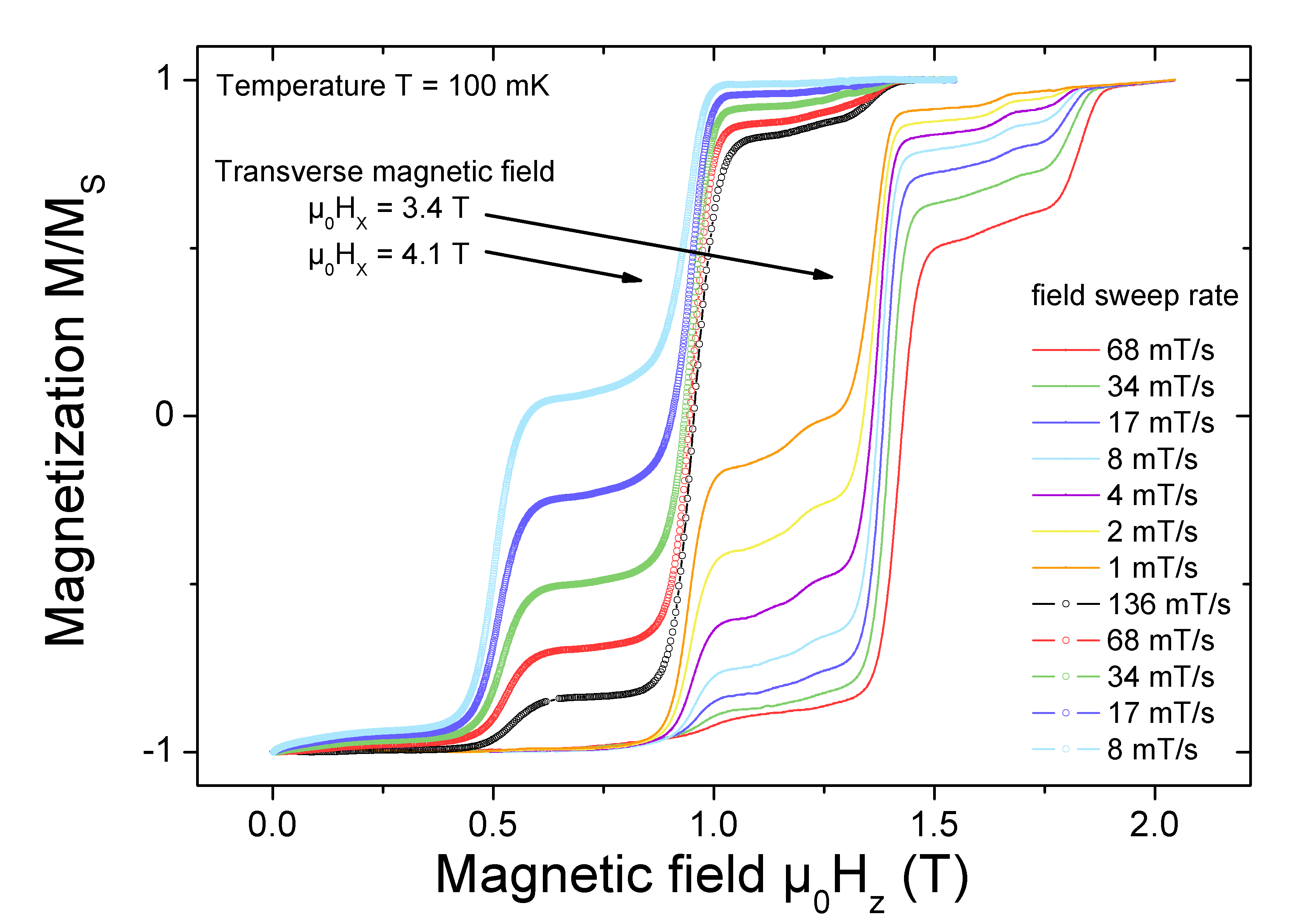}
\caption{\label{fig2} (color online)
Magnetization measurements for different field sweep rates and two transverse 
magnetic fields $\mu_0 H_x =3.4$~T (solid lines) and $\mu_0 H_x = 4.1$~T (open symbols).
The sample was first saturated in a large negative magnetic field and then 
ramped at constant sweep rate to positive field.
All measurements were done at low temperature $T=100$~mK.
}
\end{figure}

\begin{figure}[t]
\includegraphics[width=3.4 in]{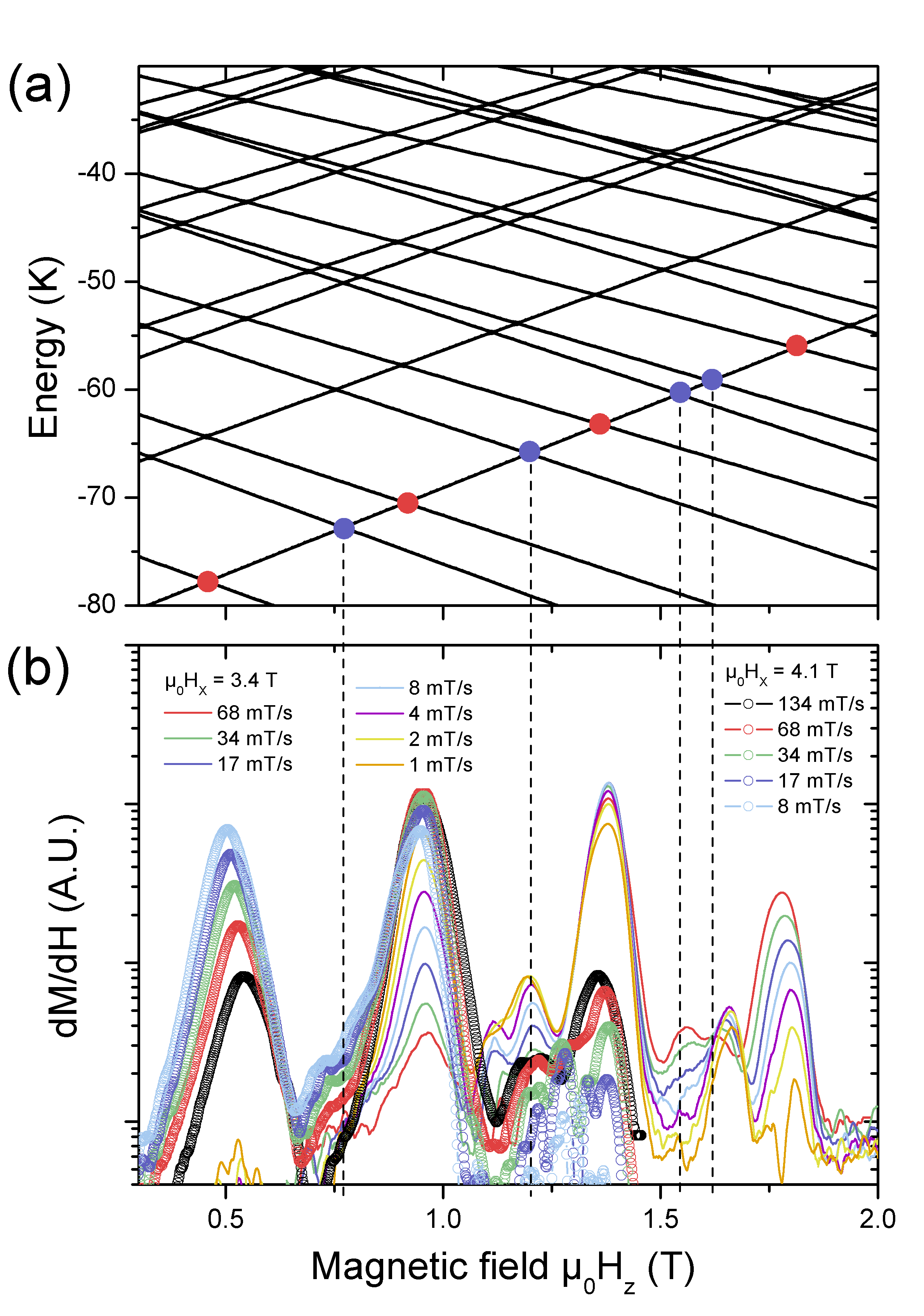}
\caption{\label{fig3} (color online)
(a) Zeeman diagram with tunnel transitions between symmetric spin states within the ground state multiplet and tunnel transitions from the ground state multiplet to spin states belonging to excited spin multiplets.
(b) The derivative plot of the magnetization curves of figure 2 shows various peaks due to tunnel transitions.
In between the main tunnel transitions between states belonging to the ground state multiplet we observe a fine structure of additional tunnel transitions involving excited spin multiplets. 
}
\end{figure}

The magnetic measurements on a single, $\mu$m-sized crystal were carried out in a dilution refrigerator employing a vector magnet system and a Hall sensor \cite{petukhov:2007}.
The easy axis of the magnetization of the crystal was parallel to the z-direction of the applied magnetic field.
The tunnel splittings of the anticrossings were determined following the Landau-Zener technique \cite{wernsdorfer:science1999,wernsdorfer:epl2000,wernsdorfer:2000,wernsdorfer:2002b,wernsdorfer:prl2006mn12}.

Figure 2 shows some magnetization measurements at low temperature for large transverse magnetic fields at different field sweep rates.
Equally spaced and very pronounced steps of the magnetization appear at approximately $\mu_0 H_z \approx 0$, $0.45$, $0.9$, $1.35$ and $1.8$~T.
In between these tunnel transitions, we observe a fine structure of smaller steps, which occur at approximately $\mu_0 H_z \approx 1.2$ and $1.65$~T.

Figure 3 shows the derivatives of the magnetization curves of figure 2 as well as the corresponding Zeeman diagram with the lowest energy levels.
The main steps of magnetization, equally spaced by $\Delta \mu_0 H_z \approx 0.45$~T, can be explained in the framework of a giant spin approximation, when describing the molecule by a collective spin $S=12$. 
We checked that the fine structure is not due to spin-spin cross-relaxation \cite{wernsdorfer:2002b}.
However, the fine structure in the magnetization steps is related to excited spin multiplets.
These steps can be understood when considering excited spin multiplets in the multi-spin approach.
The tunnel transition at $\mu_0 H_z \approx 0.45$~T involves the symmetric eigenstates $\ket{-6,-6}$ and $\frac{1}{\sqrt{2}}(\ket{6, 5}+\ket{5, 6})$.
In fact, in between the main, equally spaced tunnel transitions several avoided level crossings appear involving excited spin multiplets (as shown by the blue dots in figure 3a).
As an example, the avoided level crossing at \mbox{$\mu_0 H_z \approx 0.75$~T} involves the symmetric eigenstate $\ket{-6,-6}$ and the antisymmetric eigenstate $\frac{1}{\sqrt{2}}(\ket{6, 5}-\ket{5, 6})$.
The tunnel process at \mbox{$\mu_0 H_z \approx 1.2$~T} involves the symmetric eigenstate $\ket{-6,-6}$ and the antisymmetric eigenstate $\frac{1}{\sqrt{2}}(\ket{ 6,  4}-\ket{ 4, 6})$.
The observed avoided level crossings in our experiments allow us to determine the longitudinal anisotropy parameter \mbox{$D=1.28$~K} and the isotropic exchange constant \mbox{$J=0.8$~K}.

\begin{figure}
\includegraphics[width=3.4 in]{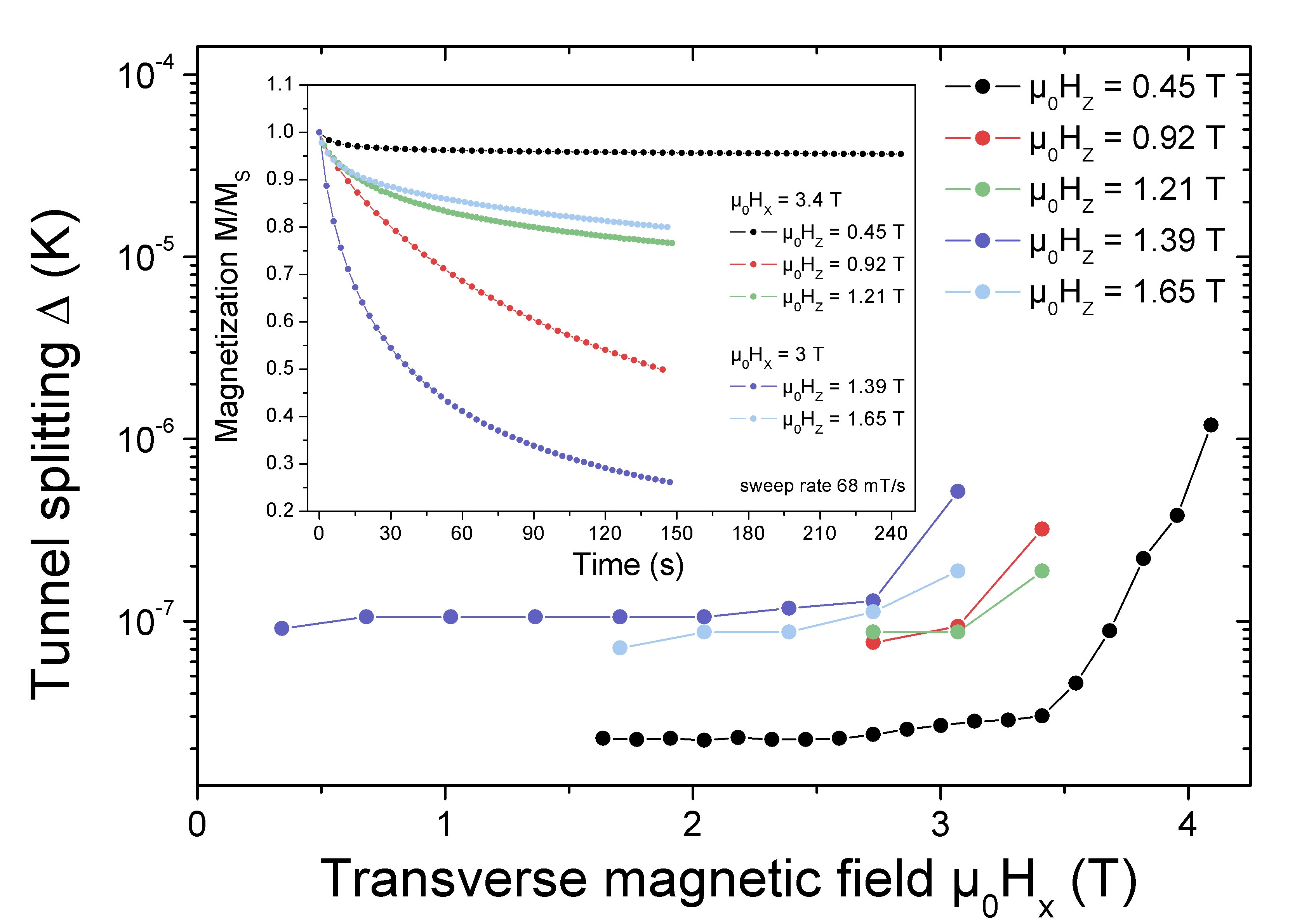}
\caption{\label{fig4} (color online)
Tunnel splittings $\Delta$ as a function of the transverse magnetic field for different level anticrossings.
The longitudinal magnetic field $\mu_0 H_z$ was swept over an avoided level crossing at a constant sweep rate $dH_z/dt=68$~mT/s and with fixed transverse magnetic field. 
The tunnel splittings were obtained by applying the Landau-Zener formula.
The inset shows the time dependence of the magnetization in the saturated sample when sweeping several times over the level anticrossing. 
}
\end{figure}

Figure 4 shows the tunnel splittings $\Delta$ of the different level anticrossings within the ground state multiplet (at $\mu_0 H_z \approx 0.45$, $0.9$ and $1.35$~T) and the ones involving excited spin multiplets (at $\mu_0 H_z \approx 1.2$ and $1.65$~T) as a function of the transverse magnetic field $\mu_0 H_x$. 
Note, that the tunnel splittings of the two antisymmetric level anticrossings around $\mu_0 H_z \approx 1.6$~T could not be studied seperately as they are too close.
In order to determine the tunnel splittings the longitudinal magnetic field was swept over a level anticrossing with fixed sweep rate $\frac{dH_z}{dt}=68\frac{mT}{s}$ and fixed transverse magnetic field $\mu_0 H_x$ and the probability of tunneling from one state to the other was measured by means of the magnetization decrease of the saturated sample.
The tunnel probability $P_{m,m'}$ between two spin states $m$ and $m'$ is given by the Landau-Zener formula
\begin{eqnarray}
	P_{m,m'} = 1- \exp \left(-\frac{\pi \Delta_{m,m'}^2}{2 \hbar g \mu_{\rm B}\mid m-m'\mid \mu_0 dH_z/dt} \right) \nonumber
\end{eqnarray}
which allows calculation of the tunnel splitting of the avoided level crossing $\Delta_{m,m'}$ when $P_{m,m'} \ll 1$ \cite{wernsdorfer:science1999,wernsdorfer:epl2000,wernsdorfer:2000,wernsdorfer:2002b,wernsdorfer:prl2006mn12}. 
Note, that $P_{m,m'} \ll 1$ is not fulfilled for very high transverse magnetic fields and therefore the experimentally obtained tunnel splittings are only estimates of a lower bound of $\Delta_{m,m'}$.
The experimentally obtained tunnel splittings lie in the range of $10^{-7}$~K for all the observed transitions and they rapidly increase when applying a transverse magnetic fields $\mu_0 H_x > 3$~T.

We found that the tunnel splittings of the anticrossings between symmetric states are mainly determined by the symmetric spin operators such as the second and fourth order anisotropy terms or the Heisenberg exchange interaction.
However the splitting between a symmetric and an antisymmetric spin state is given by the matrix element involving antisymmetric spin operators, i.e. in the framework of our model the antisymmetric Dzyaloshinskii-Moriya interaction.
When analyzing the tunnel splitting between a symmetric and an antisymmetric spin state we can get an estimate of the Dzyaloshinskii-Moriya interaction parameter $\vec{D}_{12}$.
Further on, the magnitude of the tunnel splitting between symmetric spin states can be used to fix the parameters D, J and possible fourth order parameters.

Numerical simulations of the tunnel splittings by exact diagonalization of the above defined Hamiltonian show, that the isotropic exchange and weak higher order transverse anisotropy terms together with a transverse magnetic field comparable to the one used in the experiments give rise to tunnel splittings between states of the groundstate multiplet on the order of $10^{-7}$~K.
The magnitude of the tunnel splittings is well reproduced when introducing a weak fourth order spin operator term as proposed by \emph{Carretta et al.} \cite{carretta:prl2008}
The strong increase in the tunnel splitting $\Delta_{m,m'}$ for large transverse magnetic fields is also well reproduced in the framework of this model.

The large tunnel splittings between symmetric and antisymmetric spin states cannot be reproduced by any symmetric spin operator such as second and fourth order crystal field anisotropy terms or the Heisenberg exchange interaction. 
However, the antisymmetric Dzyaloshinskii-Moriya exchange interaction can provide the quite large coupling between the symmetric and antisymmetric spin states.
Numerical simulations show that a vector of the Dzyaloshinskii-Moriya interaction $\vec{D}_{12}$ with components $D_x=D_y=D_z=10$~mK gives rise to tunnel splittings of the order of the experimentally observed ones.
In particular, the calculated tunnel splitting at $\mu_0 H_z \approx 0.75$~T turns out to be at least one order of magnitude smaller than the one at $\mu_0 H_z \approx 1.2$~T. 
This is consistent with our experiments, as we did not observe any clear and pronounced step in the magnetization curves at $\mu_0 H_z \approx 0.75$~T.
The corresponding tunnel splitting is theoreticaly - with the parameters given above and transverse magnetic fields below 4~T - smaller than $10^{-8}$~K and therefore too small to be measured with our experimental technique.

In conclusion, we presented magnetization measurements on the SMM $\rm Mn_6$ revealing various tunnel transitions, which are forbidden in the framework of a giant spin approximation. 
We propose to describe the $\rm Mn_6$ SMM as a molecular dimer of two coupled spins $S=6$. 
The introduction of an antisymmetric exchange interaction leads to the superposition of spin states with different spin length. 
This superposition of spin states belonging to different multiplets leads to additional tunnel transitions which are observed in our experiments and are in perfect agreement to our
theoretical model.
This multi-spin description goes far beyond the standard giant spin approximation and is capable of explaining the experimentally observed tunnel transitions.
This dimeric model of the molecule is confirmed by numerical calculations of the positions and the magnitude of the tunnel splittings which are consistent with the experimental results.

\begin{acknowledgments}
This work is partially financed by EC-RTN-QUEMOLNA Contract No.
MRTN-CT-2003-504880 and MAGMANet. 
We thank E. Eyraud and D. Lepoittevin for technical support.
\end{acknowledgments}


\begin{thebibliography}{10}

\bibitem{novak:1995}
M. Novak and R. Sessoli,  in {\em Quantum Tunneling of Magnetization-QTM'94
  (NATO ASI Series E : Applied Sciences)}, edited by L. Gunther and B. Barbara
  (Kluwer, Dordrecht, 1995), Vol.~301, pp.\ 171--188.

\bibitem{friedman:1996}
J.~R. Friedman, M.~P. Sarachik, J. Tejada, and R. Ziolo, Phys. Rev. Lett. {\bf
  76},  3830  (1996).

\bibitem{thomas:1996}
L. Thomas, F. Lionti, R. Ballou, D. Gatteschi, R. Sessoli, and B. Barbara,
  Nature {\bf 383},  145  (1996).

\bibitem{gatteschi:2003}
D. Gatteschi and R. Sessoli, Angew. Chem. Int. Ed. {\bf 42},  268  (2003).

\bibitem{wernsdorfer:science1999}
W. Wernsdorfer and R. Sessoli, Science {\bf 284},  133  (1999).

\bibitem{leuenberger:2001}
M.~N. Leuenberger and D. Loss, Nature {\bf 410},  789  (2001).

\bibitem{troiani:2005}
F. Troiani, A. Ghirri, M. Affronte, S. Carretta, P. Santini, G. Amoretti, S.
  Piligkos, G.~A. Timco, and R.~E.~P. Winpenny, Phys. Rev. Lett. {\bf 94},
  207208  (2005).

\bibitem{ardavan:2007}
A. Ardavan, O. Rival, J.~J.~L. Morton, S.~J. Blundell, A.~M. Tyryshkin, G.~A.
  Timco, and R.~E.~P. Winpenny, Phys. Rev. Lett. {\bf 98},  057201  (2007).

\bibitem{caneschi:jacs1991}
A. Caneschi, D. Gatteschi, R. Sessoli, A.-L. Barra, L.~C. Brunel, and M.
  Guillot, Journal of the American Chemical Society {\bf 113},  5873  (1991).

\bibitem{sessoli:jacs1993}
R. Sessoli, H.~L. Tsai, A.~R. Schake, S. Wang, J.~B. Vincent, K. Folting, D.
  Gatteschi, G. Christou, and D.~N. Hendrickson, Journal of the American
  Chemical Society {\bf 115},  1804  (1993).

\bibitem{barra:1996}
A.-L. Barra, P. Debrunner, D. Gatteschi, Ch.~E. Schulz, and R. Sessoli,
  Europhys. Lett. {\bf 35},  133  (1996).

\bibitem{petukhov:2004}
K. Petukhov, S. Hill, N.~E. Chakov, K.~A. Abboud, and G. Christou, Phys. Rev. B
  {\bf 70},  054426  (2004).

\bibitem{barco:2004}
E. del Barco, A.~D. Kent, E.~C. Yang, and D.~N. Hendrickson, Phys. Rev. Lett.
  {\bf 93},  157202  (2004).

\bibitem{carretta:2007prl}
S. Carretta, P. Santini, G. Amoretti, T. Guidi, J.~R.~D. Copley, Y. Qiu, R.
  Caciuffo, G. Timco, and R.~E.~P. Winpenny, Phys. Rev. Lett. {\bf 98},  167401
   (2007).

\bibitem{barco:nature2008}
C.M. Ramsey, E. del Barco, S. Hill, S.J. Shah,
  C.C. Beedle, and D.N. Hendrickson, Nature Physics {\bf 4},  277  (2008).

\bibitem{carretta:prl2008}
S. Carretta, T. Guidi, P. Santini, G. Amoretti, O. Pieper, B. Lake, J. van
  Slageren, F.~El Hallak, W. Wernsdorfer, H. Mutka, M. Russina, C.~J. Milios,
  and E.~K. Brechin, Phys. Rev. Lett. {\bf 100},  157203  (2008).

\bibitem{Milios_JACS2}
C.J. Milios, A. Vinslava, W. Wernsdorfer, S. Moggach, S. Parsons, S.P.
  Perlepes, G. Christou, and E.K. Brechin, Journal of the American Chemical
  Society {\bf 129},  2754  (2007).

\bibitem{petukhov:2007}
K. Petukhov, S. Bahr, W. Wernsdorfer, A.-L. Barra, and V. Mosser, Phys. Rev. B
  {\bf 75},  064408  (2007).

\bibitem{wernsdorfer:2000}
W. Wernsdorfer, R. Sessoli, A. Caneschi, D. Gatteschi, A. Cornia, and D.
  Mailly, J. Appl. Phys. {\bf 87},  5481  (2000).

\bibitem{wernsdorfer:2002b}
W. Wernsdorfer, S. Bhaduri, R. Tiron, D.~N. Hendrickson, and G. Christou, Phys.
  Rev. Lett. {\bf 89},  197201  (2002).

\bibitem{wernsdorfer:prl2006mn12}
W. Wernsdorfer, M. Murugesu, and G. Christou, Phys. Rev. Lett. {\bf 96},
  057208  (2006).

\bibitem{wernsdorfer:epl2000}
W. Wernsdorfer, R. Sessoli, A. Caneschi, D. Gatteschi, and A. Cornia, Europhys.
  Lett. {\bf 50},  552  (2000).

\end{thebibliography}
\end{document}